\documentclass[sn-chicago]{sn-jnl}% Chicago-based Humanities Reference Style

\usepackage{graphicx}%
\usepackage{multirow}%
\usepackage{amsmath,amssymb,amsfonts}%
\usepackage{amsthm}%
\usepackage{mathrsfs}%
\usepackage[title]{appendix}%
\usepackage{xcolor}%
\usepackage{textcomp}%
\usepackage{manyfoot}%
\usepackage{booktabs}%
\usepackage{algorithm}%
\usepackage{algorithmicx}%
\usepackage{algpseudocode}%
\usepackage{listings}%
\usepackage{pifont}
\newcommand{\cmark}{\ding{51}}%
\newcommand{\xmark}{\ding{55}}%

\def\ep{E}
\def\var{Var}
\def\cov{Cov}

\theoremstyle{thmstyleone}%
\newtheorem{remark}{Remark}%

\newtheorem{lemma}{Lemma}[section]

\begin{document}

\title[A fast and effective kernel two-sample test for large-scale data]{A fast and effective kernel two-sample test for large-scale data}

\author[1]{\fnm{Hoseung} \sur{Song}}\email{hoseung@kaist.ac.kr}

\author*[2]{\fnm{Hao} \sur{Chen}}\email{hxchen@ucdavis.edu}

\affil[1]{\orgdiv{Department of Industrial and Systems Engineering}, \orgname{KAIST}, \orgaddress{\city{Daejeon}, \postcode{34141}, \country{Republic of Korea}}}

\affil*[2]{\orgdiv{Department of Statistics}, \orgname{University of California, Davis}, \orgaddress{\street{One Shields Avenue}, \city{Davis}, \postcode{95616}, \state{CA}, \country{USA}}}

%%%%%%%%%%%%%%%%%%%%%%%%%%%%%%%%%%%%%%%%%%%%%%%%%%%%%%%%%%%%%%%%%%%%%%%%%%%%%%

\abstract{Kernel two-sample tests have been widely used, and the development of efficient methods for high-dimensional, large-scale data is receiving increasing attention in the big data era. However, existing methods, such as the maximum mean discrepancy (MMD) and recently proposed kernel-based tests for large-scale data, are computationally intensive and/or ineffective for some common alternatives in high-dimensional data. In this paper, we propose a new test that exhibits high power across a wide range of alternatives. Furthermore, the new test is more robust to high dimensions than existing methods and does not require optimization procedures for choosing kernel bandwidth and other parameters through data splitting. Numerical studies demonstrate that the new approach performs well on both synthetic and real-world data.}


%%%%%%%%%%%%%%%%%%%%%%%%%%%%%%%%%%%%%%%%%%%%%%%%%%%%%%%%%%%%%%%%%%%%%%%%%%%%%%

\allowdisplaybreaks

\maketitle

%%%%%%%%%%%%%%%%%%%%%%%%%%%%%%%%%%%%%%%%%%%%%%%%%%%%%%%%%%%%%%%%%%%%%%%%%%%%%%

\section{Introduction}

Two-sample hypothesis testing plays a significant role in a variety of scientific applications, such as bioinformatics, social sciences, and image analysis \citep{fox2006two, osborne2013nonparametric, kohout2017network}. As we enter the big data era, high-dimensional and large-scale data is becoming prevalent, particularly in machine learning and deep learning applications. Consequently,  attention to  two-sample testing methods for large-scale data is  naturally increasing \citep{sutherland2016generative, grosse2017statistical, carlini2017adversarial, gao2020maximum}.

Formally, given two samples $X_{1}, X_{2}, \ldots, X_{m} \stackrel{iid}{\sim} P$ and $Y_{1}, Y_{2},  \ldots, Y_{n} \stackrel{iid}{\sim} Q$, where $P$ and $Q$ are distributions in $\mathcal{R}^{d}$, we consider testing the null hypothesis $H_{0}: P = Q$ against $H_{1}: P \ne Q$. Classical approaches have focused on univariate data \citep{kolmogorov1933sulla, wald1940test, mann1947test}. In the multivariate setting, there are many parametric methods \citep{bai1996effect, schott2007test, srivastava2008test, srivastava2010testing}, but they are limited by model assumptions. Nonparametric approaches are more flexible, and several tests have been proposed utilizing the rank \citep{baumgartner1998nonparametric,hettmansperger1998affine,rousson2002distribution,oja2010multivariate}, interpoint distances \citep{szekely2013energy, biswas2014nonparametric}, graphs \citep{friedman1979multivariate, schilling1986multivariate, rosenbaum2005exact, chen2013graph, chen2017new, chen2018weighted, zhang2017graph}, ball divergence \citep{pan2018ball}, and classifiers \citep{lopez2016revisiting, hediger2019use, liu2020learning, kirchler2020two}. In this paper, we focus on kernel two-sample tests, and some of the aforementioned state-of-the-art tests will be included for comparison in numerical studies.

%%%%%%%%%%%%%%%%%%%%%%%%%%

\subsection{A brief review of kernel two-sample tests}

As a nonparametric framework, kernel two-sample tests have been widely used in real data applications \citep{liu2020learning, wynne2020kernel}. The most well-known kernel two-sample test was proposed by \cite{gretton2007kernel}. Given a kernel $k(\cdot,\cdot)$, the test is based on the maximum mean discrepancy (MMD), which measures the difference between expected features of $P$ and $Q$ in a reproducing kernel Hilbert space (RKHS):
\begin{align*}
	\textrm{MMD}^2(P,Q) = E_{X,X'}[k(X,X')] - 2E_{X,Y}[k(X,Y)] + E_{Y,Y'}[k(Y,Y')],
\end{align*}
where $X$ and $X'$ are independent random variables drawn from $P$, and $Y$ and $Y'$ are independent random variables drawn from $Q$. 

An unbiased estimate of $\textrm{MMD}^2$ was considered by \cite{gretton2007kernel}: 
\begin{align*}
	\textrm{MMD}^2_{u} &= \frac{1}{m(m-1)}\sum_{i=1}^{m}\sum_{j=1,j\ne i}^{m}k(X_{i},X_{j}) + \frac{1}{n(n-1)}\sum_{i=1}^{n}\sum_{j=1,j\ne i}^{n}k(Y_{i},Y_{j})  \\
	&\ \ \ \ \ \ \ - \frac{2}{mn}\sum_{i=1}^{m}\sum_{j=1}^{n}k(X_{i},Y_{j}) \\
	&\stackrel{\triangle}{=} \alpha + \beta - 2\gamma.
\end{align*}
Here, $\alpha$ and $\beta$ represent the average of within-sample kernel values, and $\gamma$ represents the average of between-sample kernel values. When the kernel $k$ is characteristic, such as the Gaussian kernel or the Laplacian kernel, the MMD behaves as a metric \citep{sriperumbudur2010hilbert}.

\cite{gretton2007kernel, gretton2009fast, gretton2012kernel} proposed testing methods based on the asymptotic behaviors of MMD as well as the bootstrap procedure. \cite{gretton2012optimal} and \cite{ramdas2015adaptivity} studied the choice of the kernel and its bandwidth, revealing that under some conditions, the performance of the test using a Gaussian kernel is independent of the bandwidth when the bandwidth is greater than the median heuristic\footnote{The median of all pairwise distances among observations.}. 

%%%%%%%%%%%%%%%%%%%%%%%%%%

\subsection{Kernel two-sample tests for large-scale data}

Although MMD works performs well in some settings, it costs $O(N^2)$ to compute the kernel values for the $N=m+n$ samples and it is computationally intensive when $N$ is large. This quadratic cost of pairwise kernel evaluations renders the classical MMD test impractical in terms of computation time or memory. For example, on a 16GB machine, storing the dense double-precision $N\times N$ kernel matrix alone exceeds available memory once $N \ge 4.5\times 10^4$. Throughout this paper, we refer to a problem as large-scale by either the $O(N^2)$ cost exceeds our compute budget or the pairwise matrix does not fit in memory. Such large-scale scenarios are increasingly common in modern two-sample testing problems, including the comparison of deep learning embeddings \citep{sutherland2018demystifying, jayasumana2024rethinking} and the analysis of single-cell or imaging data \citep{ozier2024kernel, meng2023lsmmd}.

A straightforward approach to handling such large-scale data is to apply subsampling, whereby the test is performed on a randomly selected subset of the full samples. While this strategy can substantially reduce computational time, it inevitably discards part of the available signal, which may lead to a noticeable loss in testing power. Consider a simple example involving multivariate log-normal data: $X_{1}, \ldots, X_{m} \stackrel{iid}{\sim} \exp(N_{d}(\textbf{0}_{d},\Sigma))$ and $Y_{1}, \ldots, Y_{n} \stackrel{iid}{\sim} \exp(N_{d}(a\textbf{1}_{d},\Sigma))$, where $\Sigma_{(i,j)} = 0.4^{|i-j|}$, $a=0.03$, and $d=100$. Table \ref{tab:subsampling} reports the estimated power of the MMD tests based on 100 simulation runs, along with the results under 50\% subsampling. For the subsampling case, within each original replication, 100 subsamples of size 1000 (50\% of the original data) are drawn and the mean power across these subsamples is reported. The results indicate that even with 50\% subsampling, the power decreases substantially. This limitation becomes more pronounced in large-scale settings, where the computational cost remains essentially of the same order as that of the full-sample analysis.
\begin{table}[h!]
	\caption{\label{tab:subsampling}Estimated power of MMD tests at 0.05 significance level with 50\% subsampling}
	\centering
	\begin{tabular}{c|c}
		\hline
		$m=n=2000$ & 50\% subsampling $(m=n=1000)$  \\
		\hline
		0.95 & 0.688 \\
		\hline
	\end{tabular}
\end{table}

%%%%%%%%%%%%%%%%%%%%%%%%%%

\subsection{Fast kernel two-sample tests and their limitations}

Since the time complexity of MMD-based tests is problematic for large-scale data, several solutions have been proposed to address this, however, all these approaches have some drawbacks and they are still time-consuming to implement (Section \ref{subsec:mmdbased} and \ref{subsec:freq} below).

%%%%%%%%%%%%%%%%%%%%%%%

\subsubsection{MMD-based work} \label{subsec:mmdbased}

\cite{gretton2012kernel} proposed a linear version of MMD, which is a running average over independent pairs of two samples. This approach is computationally more efficient than MMD, and its null distribution is straightforward to obtain. However, it tends to have low power because the variance of the test statistic is large due to the sparse division.

To balance  test power and  computational cost, \cite{zaremba2013b} considered a block averaging version of MMD.  This version achieves higher power than the linear version of MMD with only a small increase in computation complexity. However, %it performs poorly when the change is in the variance of distribution. Furthermore, 
the median heuristic for choosing the bandwidth is ineffective because the median cannot capture the main data variation by perturbations \citep{gretton2012optimal}. Finding a suitable kernel bandwidth to ensure the test's performance can be time consuming.

Additionally, both tests only work for the balanced sample design, i.e, when the sample sizes of the two samples are the same.

%%%%%%%%%%%%%%%%%%%%%%%

\subsubsection{Tests based on differences in expectations at spatial or frequency locations} \label{subsec:freq}

Recently, \cite{chwialkowski2015fast} developed two tests with a cost linear in the sample size. They demonstrated that the distance between mean embeddings in RKHS can be evaluated sufficiently at a finite number of randomly chosen test locations. Based on the differences in the analytic representations of distributions, they proposed two linear time two-sample tests: the ME test and the SCF test. The ME test uses the test statistic defined as $n\bar{z}^{T}S^{-1}\bar{z}$ when $m=n$, where $\bar{z} = \frac{1}{n}\sum_{i=1}^{n}z_{i}$, $S = \frac{1}{n-1}\sum_{i=1}^{n}(z_{i}-\bar{z})(z_{i}-\bar{z})^{T}$, and $z_{i} = \left(k(x_{j},v_{j}) - k(y_{j},v_{j})\right)_{j=1}^{J} \in \mathcal{R}^{J}$. The test statistic depends on a set of $J$ test locations $v_{j} \in \mathcal{R}^{d}$ for $j = 1, \ldots, J$, which are arbitrarily chosen. The SCF test uses a test statistic with the same form as the ME test statistic but modifies $z_{i}$ using the Fourier transform.

Later, \cite{jitkrittum2016interpretable} extended the approaches in \cite{chwialkowski2015fast} and proposed to optimize the test locations and the kernel bandwidth by maximizing a lower bound of the test power. This approach significantly improves the performance of the test with linear complexity. However, parameters such as test locations and the kernel bandwidth need to be optimized, and these steps requires substantial time. Specifically, the data must be split into two sets to learn, which may lead to a potential loss of power due to a smaller test sample size. 

In addition, the test statistics are occasionally not well-defined since $S$ in the test statistics is not invertible despite regularization.  Moreover, all these approaches also only work for balanced sample designs.

%%%%%%%%%%%%%%%%%%%%%%%%%%

\subsection{Our contribution} 

In this paper, we develop a fast and effective kernel-based test for high-dimensional large-scale data and provide the asymptotic distribution of the new test statistic, offering easy off-the-shelf tools for large datasets. The main contributions of this paper are summarized below:
\begin{enumerate}
	\item Efficiency: The new approach eliminates the need for cumbersome procedures such as optimization for kernel bandwidth or other parameters through data splitting and grid search, making it significantly faster than existing methods.
	\item Effectiveness: The new test statistic leverages patterns caused by the curse of dimensionality, performing well across a wide range of alternatives.
	\item Flexibility: This is the first fast kernel method for large-scale data capable of handling unequal sample sizes.
\end{enumerate}

Table \ref{tab:comparison} gives a high-level comparison between the proposed method and existing benchmark methods in three aspects. The approach we introduce in this paper is the only one of the four which renders all of the desired properties. The detail results can be found in Section \ref{sec:experiment}.
\begin{table}[h]
	\label{tab:comparison}
	\caption{Comparison between the proposed method and three other methods for large-scale two-sample tests}
	\begin{tabular}{|c|c|c|c|}
		\hline
		Methods         & Efficiency & Effectiveness & Flexibility \\ \hline
		\cite{zaremba2013b} (MMD-B)&  \cmark   &   \xmark &     \xmark \\ \hline
		\cite{jitkrittum2016interpretable} method 1 (ME-full) & \xmark &    \cmark &  \xmark  \\ \hline
		\cite{jitkrittum2016interpretable} method 2 (SCF-full) &  \xmark &  \cmark &   \xmark \\ \hline
		Proposed method &  \cmark   &  \cmark &    \cmark \\ \hline
	\end{tabular}
\end{table}

The organization of the paper is as follows. The new test is proposed and discussed in Section \ref{sec:new}. Section \ref{sec:experiment} examines the performance of the new test under various simulation settings. Section \ref{sec:real} illustrate the new approach with real data applications. We discuss a few other blocking approaches along similar lines in Section \ref{sec:discussion} and conclude in Section \ref{sec:conclusion}. The codes for implementing the proposed method and reproducing simulation experiments are available at GitHub (\url{https://github.com/hoseungs/Fast_kenel_two-sample_test}).

%%%%%%%%%%%%%%%%%%%%%%%%%%%%%%%%%%%%%%%%%%%%%%%%%%%%%%%%

\section{A new test} \label{sec:new}

The new test statistic builds on the block averaging approach with MMD proposed by \cite{zaremba2013b}, which was originally introduced in \cite{ho2006two}. This approach has also been utilized in computational chemistry \citep{kent2007efficient, grossfield2009quantifying}.

To balance computational time and test performance, \cite{zaremba2013b} proposed using a block size of $\sqrt{m}$, leading to a sub-quadratic time test. However, they suggested the same block sizes for two samples, making it only applicable to  balanced sample designs or resulting in wasted observations under unbalanced sample designs. %, and this approach also leads to a potential loss of power due to wasted observations (see Section \ref{sec:discussion}). 
To enable the new test to handle unbalanced sample sizes, we allow different block sizes for the two samples. Specifically, we split the data into blocks of different sizes for $X$-sample and $Y$-sample. The number of blocks is defined as $b = \lfloor\sqrt{(m+n)/2}\rfloor$\footnote{$\lfloor$x$\rfloor$ denotes the largest integer that is no larger than x.}, and the corresponding block sizes are $\lfloor m/b\rfloor$ for the $X$-sample and $\lfloor n/b\rfloor$ for the $Y$-sample if $m$ and $n$ are divisible by $b$. If $m$ or $n$ is not divisible by $b$, we allow some blocks to have $\lfloor m/b\rfloor+1$ or $\lfloor n/b\rfloor+1$ observations to fully utilize the data. Explicitly, let $B_{1,i}$, $B_{2,i}$ be the block sizes of the $i$th block for $X$-sample and $Y$-sample, respectively, and let $r_{x} = m-b\lfloor m/b\rfloor$, $r_{y} = n-b\lfloor n/b\rfloor$. Then, for $i = 1, \ldots, b$, 
\begin{equation} \label{eqn1}
	B_{1,i} = \left\{
	\begin{tabular}{c} 
		$\lfloor m/b\rfloor$ \ \ \ \  \ \ \ for $i = 1, \ldots, b-r_{x}$, \ \ \  \ \   \\
		$\lfloor m/b\rfloor + 1$ \ \ for $i = b-r_{x}+1, \ldots, b$, 
	\end{tabular}
	\right.
\end{equation}
\begin{equation} \label{eqn2}
	B_{2,i} = \left\{
	\begin{tabular}{c} 
		$\lfloor n/b\rfloor$ \ \ \ \  \ \ \  for $i = 1, \ldots, b-r_{y}$,  \ \ \  \ \  \\
		$\lfloor n/b\rfloor + 1$ \ \ for $i = b-r_{y}+1, \ldots, b$. 
	\end{tabular}
	\right.
\end{equation}

We build on core statistics proposed in \cite{song2024generalized}, which overcome the curse of dimensionality and can cover a wider range of alternatives compared to MMD. In particular, two statistics are defined as follows:
\begin{align}
	W_{i} &= \frac{B_{1,i}}{B_{i}}\alpha_{i} + \frac{B_{2,i}}{B_{i}}\beta_{i}, \label{stat:w}\\
	D_{i} &= B_{1,i}(B_{1,i}-1)\alpha_{i} - B_{2,i}(B_{2,i}-1)\beta_{i},  \label{stat:d}
\end{align}
for $i = 1, \ldots, b$ where $B_{i} = B_{1,i}+B_{2,i}$. Here, $\alpha_{i}$ and $\beta_{i}$ are the block versions of $\alpha$ and $\beta$, respectively, computed in each block. It is expected that $W_{i}$ tends to be sensitive to location alternatives and $D_{i}$ tends to be sensitive to scale alternatives. 

We adopt the permutation distribution as the null distribution, which places $1/{B_{i} \choose B_{1,i}}$ probability on each of the ${B_{i} \choose B_{1,i}}$ choices of $B_{1,i}$ out of the total $B_{i}$ observations as the $X$- sample for each block, and similarly for each block of $Y$-sample. The approach in \cite{zaremba2013b} suffers from the choice of a suitable bandwidth in kernels, and there is no specific strategy for this. As discussed in \cite{gretton2012optimal}, the median heuristic captures the global length scale in the data, so it could fail when the global length scale of the data is large, but the length scale required to capture the difference between two distributions $P$ and $Q$ is small. An example of such a problem is the Blobs data studied in \cite{gretton2012optimal} and \cite{jitkrittum2016interpretable}. However, under the permutation null distribution, the median heuristic is still a reasonable choice (see Section \ref{subsec:simul3} for more discussions).

\begin{figure}[h!]
	\centering
	\includegraphics[width=4in]{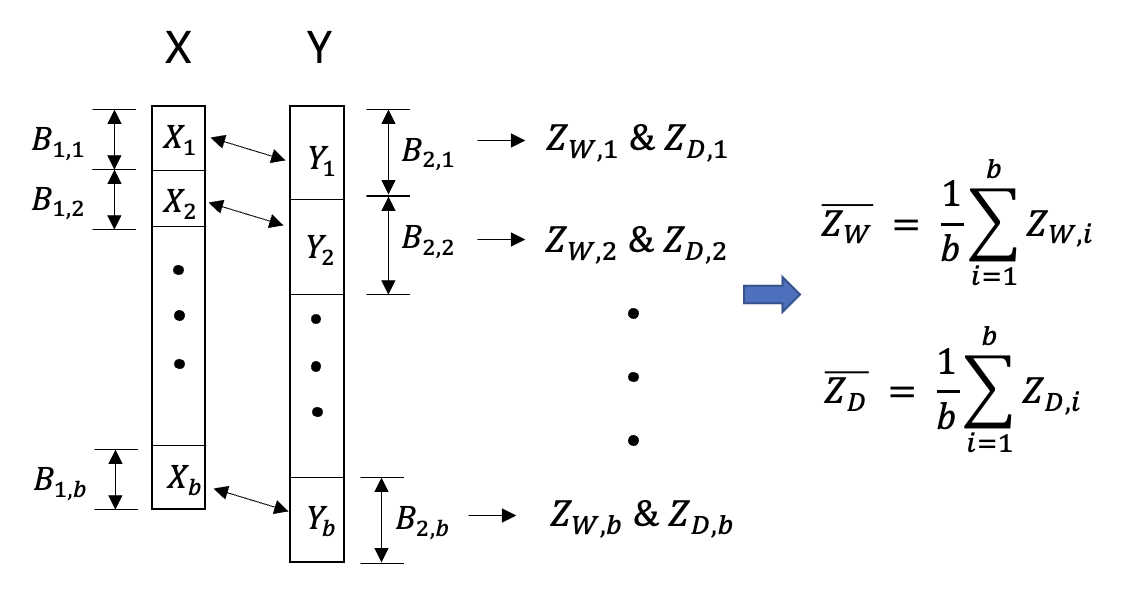}
	\caption{Illustration for the test statistics}
	\label{fig:illustration}
\end{figure}

The standardized statistics for (\ref{stat:w}) and (\ref{stat:d}) are defined as follows:
\begin{align}
	Z_{W,i} &= \frac{W_{i}-\ep\left(W_{i}\right)}{\sqrt{\var\left(W_{i}\right)}}, \label{eqn5} \\
	Z_{D,i} &= \frac{D_{i}-\ep\left(D_{i}\right)}{\sqrt{\var\left(D_{i}\right)}}, \label{eqn6}
\end{align}
for $i = 1, \ldots, b$,  where $\ep\left(W_{i}\right)$, $\var\left(W_{i}\right)$, $\ep\left(D_{i}\right)$, and $\var\left(D_{i}\right)$ are the expectations and variances of $W_{i}$ and $D_{i}$, respectively, under the permutation null distribution. Their analytic expressions can be obtained in a similar manner to those in \cite{song2024generalized} and 
are provided in Lemma \ref{lemma2.1}. Figure \ref{fig:illustration} illustrates the formulation of the new test statistic. 

For pooled observations $z_{1}^{(i)}, \ldots, z_{B_{i}}^{(i)}$ in $i$-th block $(i = 1, \ldots, b)$, $\alpha_{i}$ and $\beta_{i}$ can be rewritten as
\begin{align*}
	\alpha_{i} &= \frac{1}{B_{1,i}(B_{1,i}-1)}\sum_{u=1}^{B_{i}}\sum_{v=1,v\ne u}^{B_{i}}k(z_{u}^{(i)},z_{v}^{(i)})I_{g_{u}^{(i)}=g_{v}^{(i)}=0},\\
	\beta_{i} &=  \frac{1}{B_{2,i}(B_{2,i}-1)}\sum_{u=1}^{B_{i}}\sum_{v=1,v\ne u}^{B_{i}}k(z_{u}^{(i)},z_{v}^{(i)})I_{g_{u}^{(i)}=g_{v}^{(i)}=1},
\end{align*}
where $g_{u}^{(i)} = 0$ if observation $z_{u}^{(i)}$ is from sample $X$ and $g_{u}^{(i)} = 1$ if observation $z_{u}^{(i)}$ is from sample $Y$ in $i$-th block.
\begin{lemma} \label{lemma2.1}
	Let $k_{uv}^{(i)} = k(z_{u}^{(i)},z_{v}^{(i)})$ for $i=1,\ldots,b$. Under the permutation null, we have
	\begin{align*}
		\ep\left(\alpha_{i}\right) &= \ep\left(\beta_{i}\right) = \frac{1}{B_{i}(B_{i}-1)}R_{0}^{(i)}, \\
		\var\left(\alpha_{i}\right) &= \frac{1}{B_{1,i}^2(B_{1,i}-1)^2}\left(2R_{1}^{(i)}p_{1,i} + 4R_{2}^{(i)}p_{2,i} + R_{3}^{(i)}p_{3,i}\right) - \ep\left(\alpha_{i}\right)^2, \\
		\var\left(\beta_{i}\right) &= \frac{1}{B_{2,i}^2(B_{2,i}-1)^2}\left(2R_{1}^{(i)}q_{1,i} + 4R_{2}^{(i)}q_{2,i} + R_{3}^{(i)}q_{3,i}\right) - \ep\left(\beta_{i}\right)^2, \\
		\cov\left(\alpha_{i},\beta_{i}\right) &= \frac{R_{3}^{(i)}}{B_{i}(B_{i}-1)(B_{i}-2)(B_{i}-3)} - \ep\left(\alpha_{i}\right)\ep\left(\beta_{i}\right),
	\end{align*}
	where
	\begin{align*}
		R_{0}^{(i)} &= \sum_{u=1}^{B_{i}}\sum_{v=1,v\ne u}^{B_{i}}k_{uv}^{(i)}, \ \  \ R_{1}^{(i)} = \sum_{u=1}^{B_{i}}\sum_{v=1,v\ne u}^{B_{i}}\left(k_{uv}^{(i)}\right)^2, \\
		R_{2}^{(i)} &= \sum_{u=1}^{B_{i}}\sum_{v=1,v\ne u}^{B_{i}}\sum_{r=1,r\ne v, r\ne u}^{B_{i}}k_{uv}^{(i)}k_{ur}^{(i)}, \\ 
		R_{3}^{(i)} &= \sum_{u=1}^{B_{i}}\sum_{v=1,v\ne u}^{B_{i}}\sum_{r=1,r\ne v, r\ne u}^{B_{i}}\sum_{s=1,s\ne r, s\ne v, s\ne u}^{B_{i}}k_{uv}^{(i)}k_{rs}^{(i)},
	\end{align*}
	with
	\begin{align*}
		p_{1,i} &= \frac{B_{1,i}(B_{1,i}-1)}{B_{i}(B_{i}-1)}, \ \ p_{2,i} = \frac{B_{1,i}(B_{1,i}-1)(B_{1,i}-2)}{B_{i}(B_{i}-1)(B_{i}-2)}, \\ 
		p_{3,i} &= \frac{B_{1,i}(B_{1,i}-1)(B_{1,i}-2)(B_{1,i}-3)}{B_{i}(B_{i}-1)(B_{i}-2)(B_{i}-3)}, \\
		q_{1,i} &= \frac{B_{2,i}(B_{2,i}-1)}{B_{i}(B_{i}-1)}, \ \ q_{2,i} = \frac{B_{2,i}(B_{2,i}-1)(B_{2,i}-2)}{B_{i}(B_{i}-1)(B_{i}-2)}, \\  
		q_{3,i} &= \frac{B_{2,i}(B_{2,i}-1)(B_{2,i}-2)(B_{2,i}-3)}{B_{i}(B_{i}-1)(B_{i}-2)(B_{i}-3)}.
	\end{align*}
\end{lemma}

Given the new test statistics, the next step is to calibrate how large the test statistics need to be to provide sufficient evidence to reject the null hypothesis. The classical central limit theorem (CLT) usually can be used for the block averaging approaches as in \cite{zaremba2013b}. However, the new test statistics $Z_{W,i}$'s are independent variables with respect to $i$, but not identically distributed under the permutation null distribution.  And a similar situation $Z_{D,i}$'s. Hence, we use the Lyapunov central limit theorem to handle the situation.
We write $x_b = o(y_b)$ when $x_b$ is dominated by $y_b$ asymptotically, i.e., $\lim_{b\rightarrow\infty}(x_{b}/y_{b}) = 0$. %Let $k_{uu}^{(i)}=0$ for $u = 1,\ldots,B_{i}$ and $i=1,\ldots,b$. 
By the Lyapunov central limit theorem,  as $b\rightarrow\infty$, 
\begin{enumerate}
	\item $\sqrt{b}\bar{Z}_{W} \rightarrow N(0,1)$ when $\sum_{i=1}^{b}\ep|Z_{W,i}|^3 = o(b^{1.5})$,
	\item $\sqrt{b}\bar{Z}_{D} \rightarrow N(0,1)$ when $\sum_{i=1}^{b}\ep|Z_{D,i}|^3 = o(b^{1.5})$.
\end{enumerate}
%when $\sum_{i=1}^{b}\ep|Z_{W,i}|^3 = o(b^{1.5})$ \textrm{and}   $\sum_{i=1}^{b}\ep|Z_{D,i}|^3 = o(b^{1.5})$, as $b\rightarrow\infty$,
%\begin{align*}
%\sqrt{b}\bar{Z}_{W} &\rightarrow N(0,1), \\
%\sqrt{b}\bar{Z}_{D} &\rightarrow N(0,1),
%\end{align*}
%if
%\begin{align*}
%\sum_{i=1}^{b}\ep|Z_{W,i}|^3 = o(b^{1.5}) \ \ \textrm{and}  \ \ \sum_{i=1}^{b}\ep|Z_{D,i}|^3 = o(b^{1.5}).
%\end{align*}

%\begin{remark}
%	The classical central limit theorem usually can be used for the block averaging approaches as in \cite{zaremba2013b}. However, the new test statistics $Z_{W,i}$'s are independent variables with respect to $i$, but not identically distributed under the permutation null distribution same for $Z_{D,i}$'s. Thus, the Lyapunov central limit theorem is used to handle the dependency and the condition \ref{condition} is the Lyapunov's condition.
%\end{remark}

\begin{figure}[h!]
	\centering
	\includegraphics[width=\columnwidth]{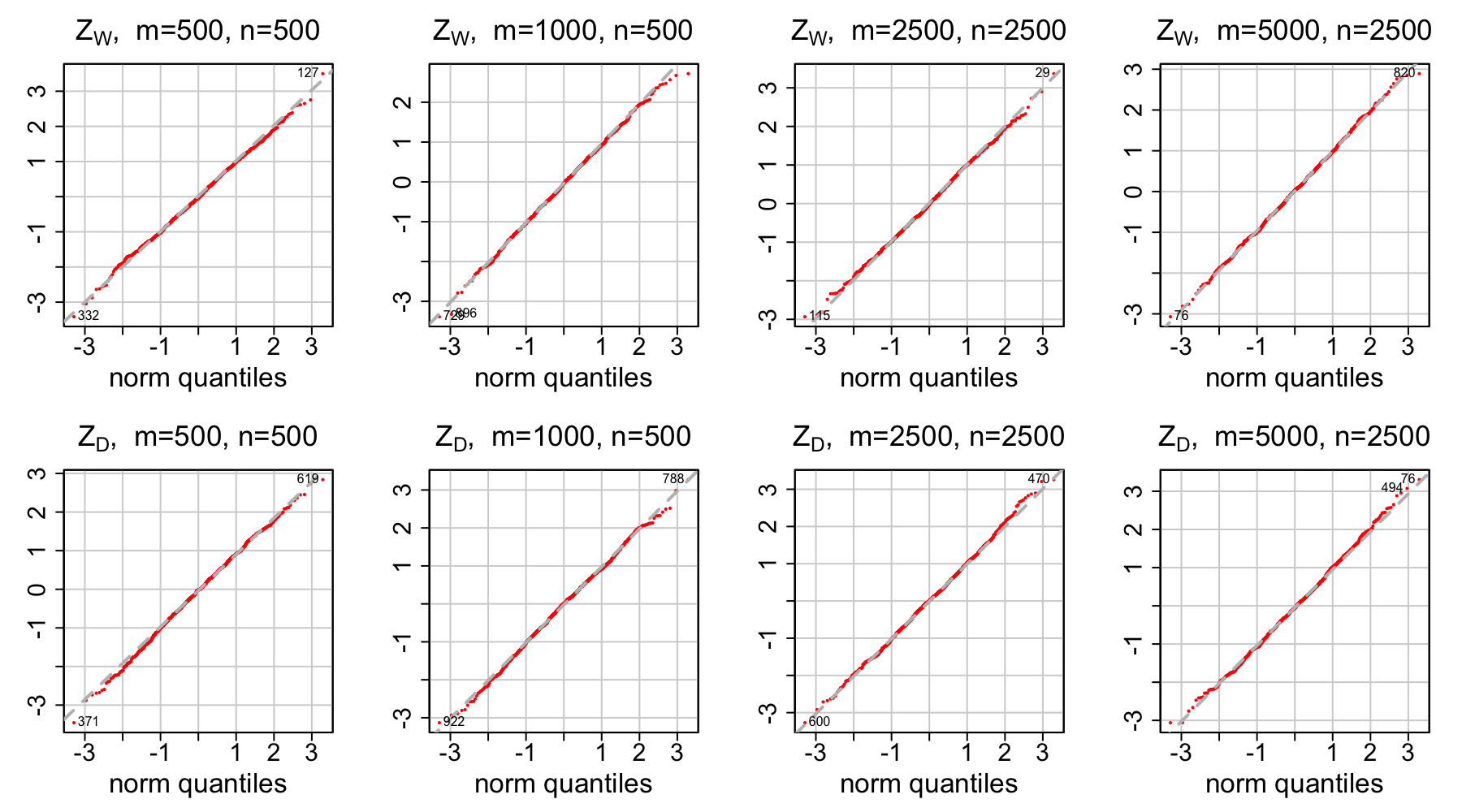}
	\caption{Normal quantile-quantile plots (red dots) of $\sqrt{b}\bar{Z}_{W}$ and $\sqrt{b}\bar{Z}_{D}$ with the gray dashed line as the baseline, which goes through the origin and has a slope of 1}
	\label{fig:qq}
\end{figure}

For the two conditions, they are satisfied if $\ep|Z_{W,i}|^3, \ep|Z_{D,i}|^3 =o(\sqrt{B_i})$, which are mild conditions since $Z_{W,i}$ and $Z_{D,i}$ are standardized statistics with mean 0 and variance 1, and they themselves are close to the normal distribution based on results in \cite{song2024generalized}. To see how the normal approximations work for finite samples, we check the normal quantile-quantile plots of $\sqrt{b}\bar{Z}_{W}$ and $\sqrt{b}\bar{Z}_{D}$. 
Figure \ref{fig:qq} presents the quantile-quantile plots of $\sqrt{b}\bar{Z}_{W}$ and $\sqrt{b}\bar{Z}_{D}$ for multivariate log-normal data $\exp\left(N_{d}(\textbf{0}_{d},I_{d})\right)$, which varying choices of $m$ and $n$ when $d=100$. We see that the asymptotic null distributions kick in for relatively small $m$ and $n$'s.

% To this end, we generate the multivariate Gaussian and log-normal data when $d=100$ (see Section \ref{subsec:simul1} and \ref{subsec:simul2}) and check the trend of $\sum_{i=1}^{b}\ep|Z_{W,i}|^3/b^{1.5}$ and $\sum_{i=1}^{b}\ep|Z_{D,i}|^3/b^{1.5}$ under different sample sizes (Figure \ref{fig:condition}). We see that quantities $\sum_{i=1}^{b}\ep|Z_{W,i}|^3/b^{1.5}$ and $\sum_{i=1}^{b}\ep|Z_{D,i}|^3/b^{1.5}$ are much smaller than 1 even when the sample size is small and they have the decreasing trend as the sample size increases.
%
%\begin{figure}[htp!]
%	\centering
%	\includegraphics[width=6.5in]{condition.png}
%	\caption{Quantities $\sum_{i=1}^{b}\ep|Z_{W,i}|^3/b^{1.5}$ and $\sum_{i=1}^{b}\ep|Z_{D,i}|^3/b^{1.5}$ under different sample sizes $m=n$.}
%	\label{fig:condition}
%\end{figure}

To combine the advantages of the two statistics based on the asymptotic results, we adopt the Bonferroni correction. Let $p_{W}$ and $p_{D}$ be the approximated $p$-values of the test that rejects for large values of $\sqrt{b}\bar{Z}_{W}$ or $\sqrt{b}|\bar{Z}_{D}|$, respectively, based on their limiting distributions. Then, the proposed test rejects the null if $2\min(p_{W}, p_{D})$ is less than the significance level. The overall testing procedure is summarized in Algorithm \ref{alg}.
\begin{algorithm}
	\caption{Testing procedure}
	\label{alg}
	\begin{algorithmic}[1]
		\Require Observations $\{X_{j}\}_{j=1,\dots,m}$, $\{Y_{j}\}_{j=1,\dots,n}$ and the significance level $\alpha$.
		\Ensure Reject the null hypothesis $H_{0}$ if $p\le\alpha$. 
		\State Calculate the number of blocks using the formula $b = \lfloor\sqrt{(m+n)/2}\rfloor$.
		\For{$i=1$ to $b$}
		\State Determine the block sizes $B_{1,i}$ and $B_{2,i}$ for $X$-sample and $Y$-sample based on (\ref{eqn1}) and (\ref{eqn2}), respectively.
		\State Compute the test statistics $Z_{W,i}$ and $Z_{D,i}$ based on (\ref{eqn5}) and (\ref{eqn6}), respectively.
		\EndFor
		\State Compute $\sqrt{b}\bar{Z}_{W}$ and $\sqrt{b}|\bar{Z}_{D}|$.
		\State Calulate $p$-values $p_{W}$ and $p_{D}$ of $\sqrt{b}\bar{Z}_{W}$ and $\sqrt{b}|\bar{Z}_{D}|$, respectively, by the standard normal distributions.
		\State Obtain the $p$-value $p = 2\min(p_{W}, p_{D})$.
	\end{algorithmic}
\end{algorithm}

\begin{remark}
	Other global testing methods, such as the Simes procedure, might also be considered. However, $\bar{Z}_{W}$ and $\bar{Z}_{D}$ are dependent, making it difficult to show the exact level of global testing for the Simes procedure. We observe that $p_{W}$ tends to be small under  location alternatives, and $p_{D}$ tends to be small under scale alternatives. Thus, the Bonferroni correction does not introduce too much conservativeness.
\end{remark}

%%%%%%%%%%%%%%%%%%%%%%%%%%%%%%%%%%%%%%%%%%%%%%%%%%%%%%%%%%%%%%%%%%%%%%%%%%%%%%%%%%%%%%%%%%%%%%%%%%%%%%%%%%%%%%%

\section{Experiments} \label{sec:experiment}

In this section, we examine the performance of the new test under diverse settings through simulations. We first focus on the existing fast kernel two-sample tests and compare them for Gaussian data using $\texttt{Python}$ (Section \ref{subsec:simul1}). In Section \ref{subsec:simul2}, we compare the new test with other state-of-the-art nonparametric two-sample tests for non-Gaussian data under $\texttt{R}$. We also briefly evaluate the performance of the new test with different choices of bandwidth in Section \ref{subsec:simul3}.

%%%%%%%%%%%%%%%%%%%%%%%%%%%%%%%%%%%%%%%

\subsection{Simulation I: comparing to state-of-the-art fast kernel two-sample tests} \label{subsec:simul1}

We compare the new test with the existing fast kernel two-sample tests proposed by \cite{zaremba2013b} and \cite{jitkrittum2016interpretable}. We denote the test proposed  in \cite{zaremba2013b} by MMD-B, and the two tests in \cite{jitkrittum2016interpretable} as ME-full and SCF-full, respectively. We also consider  Hotelling's $\textrm{T}^2$ test as a baseline. We use the Gaussian kernel for all tests. For ME-full and SCF-full, the bandwidth of the Gaussian kernel and test locations are fully optimized on a training sample. For a fair comparison, the bandwidth of the Gaussian kernel used in MMD-B is also optimized by a grid search, while the median heuristic is used in the new test. An implementation of the existing methods is available at \url{https://github.com/wittawatj/interpretable-test}, as provided in \cite{jitkrittum2016interpretable}.

We follow the simulation setup in \cite{jitkrittum2016interpretable} to compare the test power and sample size $(m=n)$ and dimension $(d)$. Additionally, we check the computational cost of the tests. The number of test locations $J$ is set to be 5 for ME-full and SCF-full, and the significance level is set to  0.01 for all  experiments. 

\begin{table}[h!] 
	\caption{\label{tab:sim}Three toy problems}
	\centering 
	\begin{tabular}{c|c|c}
		\hline
		Data & $P$ & $Q$  \\
		\hline
		GMD   & $N_{d}(\textbf{0}_{d}, I_{d})$ & $N_{d}((0.8,0,\ldots,0)^{T}, I_{d})$ \\
		GVD & $N_{d}(\textbf{0}_{d}, I_{d})$ & $N_{d}(\textbf{0}_{d}, \textrm{diag}(2,1,\ldots,1))$ \\
		NULL    & $N_{d}(\textbf{0}_{d}, I_{d})$ & $N_{d}(\textbf{0}_{d}, I_{d})$\\
		\hline
	\end{tabular}
\end{table}

\subsubsection{Test power vs. sample size $m$} 

We study how the sample size affects test powers and type I error, considering three settings in Table \ref{tab:sim}: GMD (location shift), GVD (scale shift), and NULL (no shift). For each sample size, we simulate 500 datasets.

\begin{figure}[h!]
	\centering
	\includegraphics[width=\columnwidth]{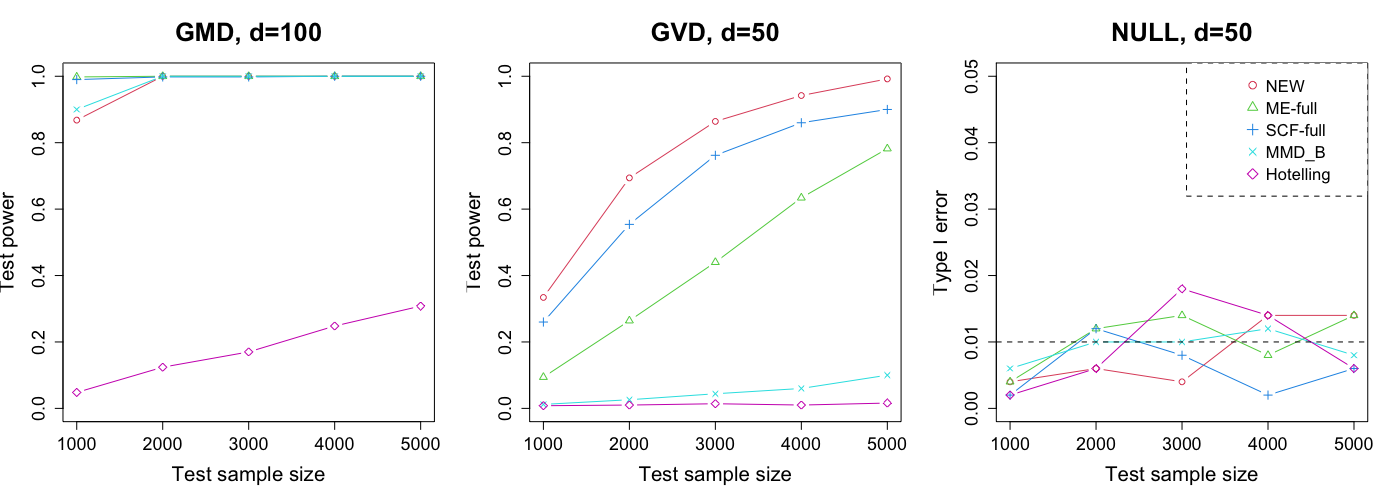}
	\caption{\label{sim:sample}Plots of test power and type I error against the test sample size $n$}
\end{figure}

The results are shown in Figure \ref{sim:sample}, where test powers (GMD and GVD) and type I error (NULL) are plotted against sample sizes. We see that the new test is comparable to the existing tests for detecting mean differences in distributions. However, the new approach achieves higher power than the existing methods for detecting  variance differences in distributions. The new test also controls the type I error rate well.

\subsubsection{Test power vs. dimension $d$} 

\begin{figure}[h!]
	\centering
	\includegraphics[width=\columnwidth]{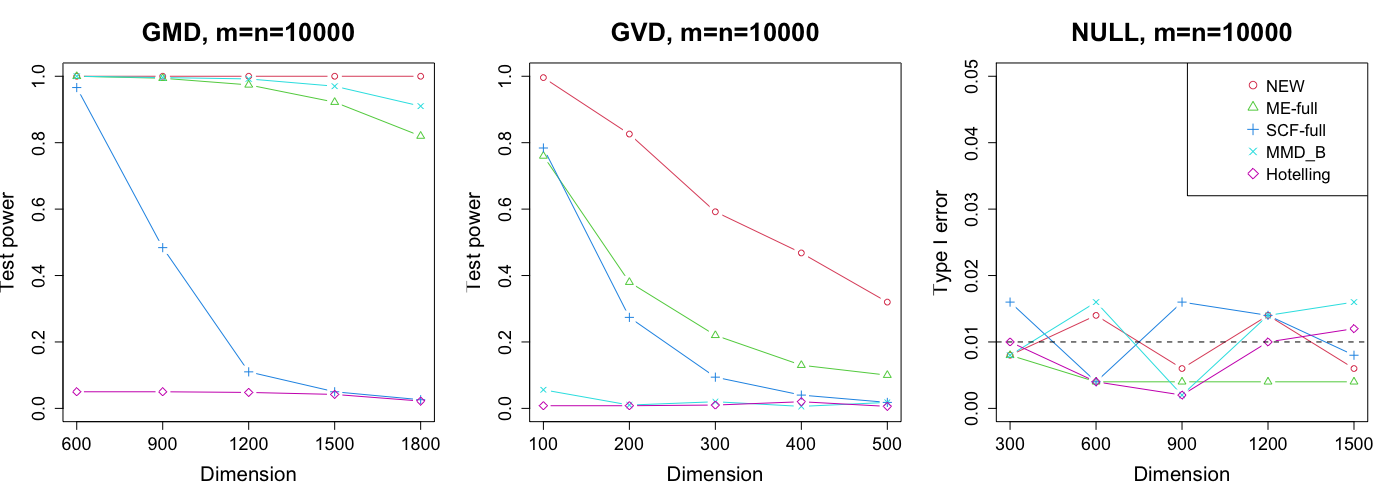}
	\caption{\label{sim:dim}Plots of test power and type I error against the dimension $d$}
\end{figure}

We next study how the dimension of the problem affects the test power and the type I error. Here, we fix the sample size to 10,000 and  simulate 500 datasets for each dimension. The results are presented in Figure \ref{sim:dim}. We observe that the new test generally dominates in power when two samples differ in either the mean or the variance. This demonstrates that the new test is powerful under high dimensions, particularly for  scale alternatives. The new test also controls the type I error rate well.

\subsubsection{Computational time} 

In addition, we compare the computational cost of the tests. One of the required properties for test on large-scale data is the computation time. We check runtimes of the tests for Gaussian data under various $m=n$. All experiments were run by $\texttt{Python}$ on 2.2 GHz Intel Core i7.

\begin{table}[h!]
	\caption{\label{tab:time_n}Average runtimes in seconds from 10 simulations for each sample size $m=n$ when $d=100$}
	\centering
	\begin{tabular}{c|cccc}
		\hline
		$m=n$ & New & ME-full & SCF-full & MMD-B  \\
		\hline
		6000    & 0.179 & 10.97 & 31.93 & 2.235 \\
		7000    & 0.194 & 12.90 & 34.86 & 2.920 \\
		8000    & 0.205 & 13.22 & 40.58 & 3.585 \\
		9000    & 0.249 & 15.03 & 49.61 &  3.805 \\
		10000    & 0.267 & 20.46 & 54.70 & 4.423 \\
		\hline
	\end{tabular}
\end{table}

Table \ref{tab:time_n} shows average runtimes for each sample size $m=n$ when $d=100$. We see that the new approach is the fastest among the existing methods. Although ME-full and SCF-full are linear time tests, they are more computationally expensive than the sub-quadratic tests MMD-B and the new test due to the parameter tuning procedures. This crucial drawback makes the existing test computationally prohibitive for large-scale data, and this problem becomes more severe as the dimension increases. 

\subsubsection{Summary} 

For the existing tests, MMD-B has low power for scale alternatives. The power of SCF-full decreases rapidly as the dimension increases. The power of ME-full decreases quickly under scale alternatives as the dimension increases. In constrst, the new test is effective for a wide range of testing problems and significantly faster for large-scale data.

%%%%%%%%%%%%%%%%%%%%%%%%%%%%%%%%%%%%%%%

\subsection{Simulation II: comparing to state-of-the-art two-sample tests} \label{subsec:simul2}

Here, we compare the new test with other state-of-the-art nonparametric two-sample tests: the graph-based test (GT) \citep{chen2017new}, the ball divergence test (BT) \citep{pan2018ball}, and the classifier two-sample test (CT) \citep{lopez2016revisiting}. We consider the following settings:
\begin{itemize}
	\item Multivariate log-normal data: $\exp(N_{d}(\textbf{0}_{d},\Sigma))$ vs.  $\exp(N_{d}(a\textbf{1}_{d},\Sigma))$. %, $\Sigma_{(i,j)} = 0.4^{|i-j|}$. %where $\Delta = \|a\textbf{1}_{d}\|_{2}$ and 
	\item Multivariate $t$-distributed data: $t_{20}(\textbf{0}_{d},\Sigma)$ vs.  $t_{20}(a\textbf{1}_{d},b\Sigma)$. %,  $\Sigma_{(i,j)} = 0.4^{|i-j|}$. %where $\Delta = \|a\textbf{1}_{d}\|_{2}$ and
	\item Chi-square data: $\Sigma^{1/2}u_1$ vs.  $(b\Sigma)^{1/2}u_2+ a\textbf{1}_{d}$, %,  $\Sigma_{(i,j)} = 0.4^{|i-j|}$, 
	where $u_1$ and $u_2$ are length-$d$ vectors with each component i.i.d. from the $\chi_{3}^2$ distribution. % where $\Delta = \|a\textbf{1}_{d}\|_{2}$,
\end{itemize}
In the above settings, $\Sigma_{(i,j)} = 0.4^{|i-j|}$.

GT, BT, and CT are publicly available in $\texttt{R}$ packages $\texttt{gTests}$, $\texttt{Ball}$, and $\texttt{Ecume}$, respectively. We simulate 500 datasets and set the significance level to 0.01 for all the experiments. 

\subsubsection{Power comparison}

\begin{figure}[h!]
	\centering
	\includegraphics[width=\columnwidth]{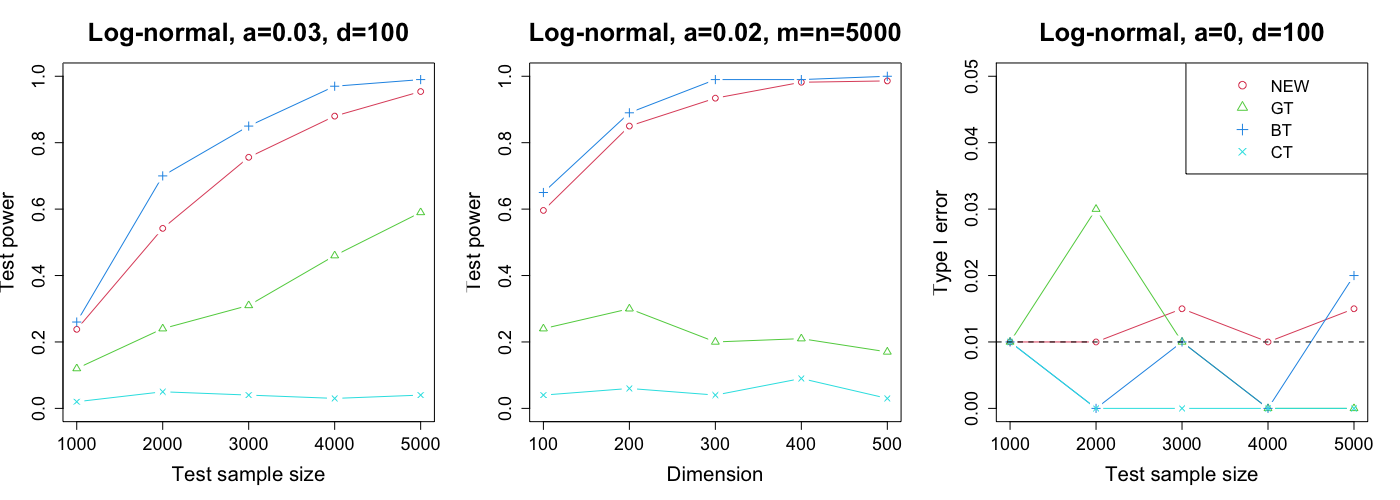}
	\includegraphics[width=\columnwidth]{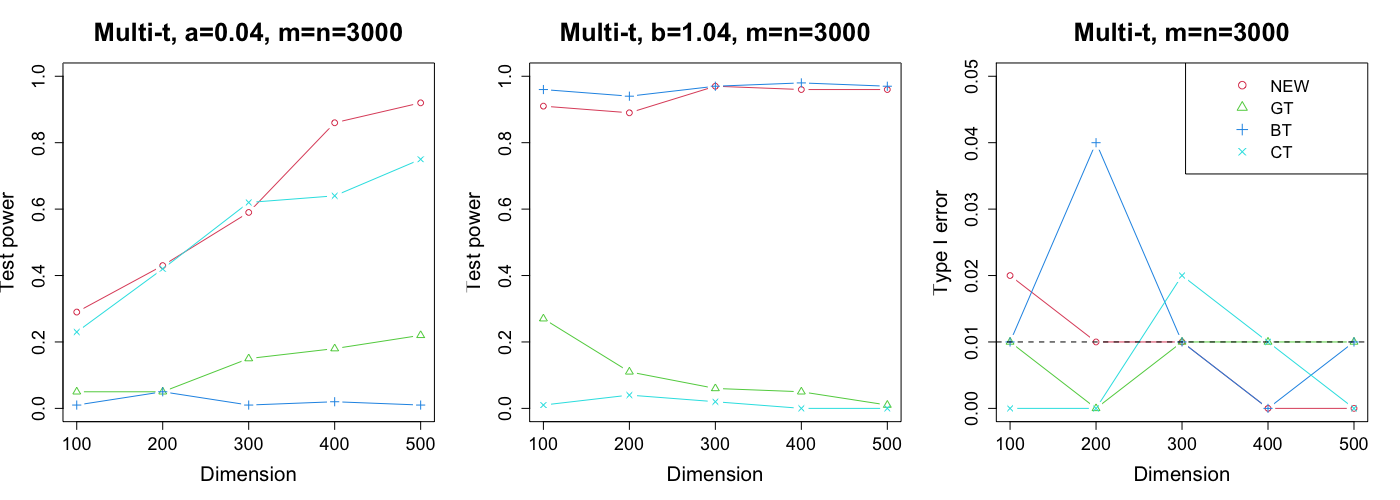}
	\includegraphics[width=\columnwidth]{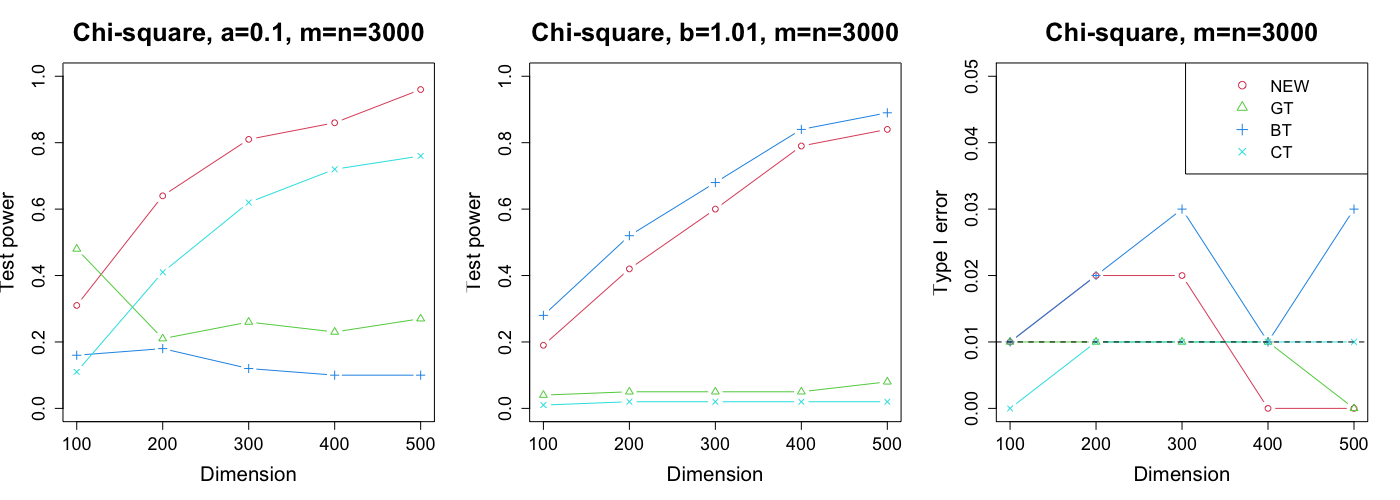}
	\caption{\label{fig:simul2}Plots of test power and type I error for the log-normal (top), multivariate $t$ (middle), and chi-square data (bottom)}
\end{figure}

We study how the sample size and the dimension affect test powers and type I error. The results are shown in Figure \ref{fig:simul2}. First, alternatives in log-normal data yield the changes in both the mean and variance of distributions. We set $a = 0.03$, $a=0.02$, and $a=0$ in the left, middle, and right top panels, respectively. For the multivariate log-normal data, we see that the new test exhibits high power for different sample sizes and dimensions,  and perform well for asymmetric distributions under moderate to high sample sizes and dimensions. GT and CT perform poorly and  are not sensitive to capturing differences in distributions. Although BT performs well, it is infeasible to use for large-scale data due to computation time (see Section \ref{subsubsec:compute}). 

The three middle and bottom panels present the results for the multivariate $t$ and chi-square data, respectively. In this case, we set the sample size to $m=n=3000$ and examine the performance under different dimensions. The left, middle, and right panels show the results for location alternatives, scale alternatives, and type I error, respectively. We see that the new test generally outperforms other tests for location alternatives and performs as well for scale alternatives. BT shows high power for scale alternatives but has no power for location alternatives. In all cases, the new test controls type I error as well.

\subsubsection{Computational time} \label{subsubsec:compute}

We compare the computation time of the tests for log-normal data using $\texttt{R}$ on 2.2 GHz Intel Core i7 under various $m=n$ (Table \ref{tab:time_another_n}). We see that the new test is the fastest among the existing tests implemented by $\texttt{R}$. Although BT shows good performance in power, it takes a long time to run and it is not useful for large-scale data. GT is faster but $\texttt{R}$ encounters memory issues storing the pairwise distances for large sample sizes.

\begin{table}[h!]
	\caption{\label{tab:time_another_n}Average runtimes in seconds from 10 simulations for each sample size $m=n$ when $d=100$}
	\centering
	\begin{tabular}{c|cccc}
		\hline
		$m=n$ & New & GT & BT & CT  \\
		\hline
		6000    & 0.485  & 80.04 & $>$2000 & 73.20 \\
		7000    & 0.609 & 106.94 & $>$2000 & 116.30 \\
		8000    & 0.748 & - & $>$2000 &200.12 \\
		9000    & 0.916 & - & $>$2000 &  165.23 \\
		10000    & 1.119 & - &  $>$2000 & 238.70 \\
		\hline
	\end{tabular}
\end{table}

%%%%%%%%%%%%%%%%%%%%%%%%%%%%%%%%%%%%%%%

\subsection{Simulation III: exploring bandwidth choice} \label{subsec:simul3} 

In this section, we briefly check whether the median heuristic is reasonable for the new test through numerical studies. We use Gaussian data $N_{d}(\textbf{0},I_{d})$ vs. $N_{d}(\mu,\sigma^2I_{d})$ with $\Delta = \|\mu\|_{2}$, and examine the average median heuristic in each setting by 100 trials. The averaged median heuristic is around 10 when $d = 100$ and 14 when $d = 200$ in our settings. So we choose 8 bandwidths that differ by 2 from each other, starting from the averaged median heuristic -8 to the averaged median heuristic +8 in order to check bandwidths in a wide range. 

We also consider the Blobs problem studied in \cite{gretton2012optimal} and \cite{jitkrittum2016interpretable}. They noted that the median heuristic may not be a good choice when the length scale required to capture the difference between two distributions $P$ and $Q$ is small, and the Blobs data has such problem.

\begin{figure}[h!]
	\centering
	\includegraphics[width=4in]{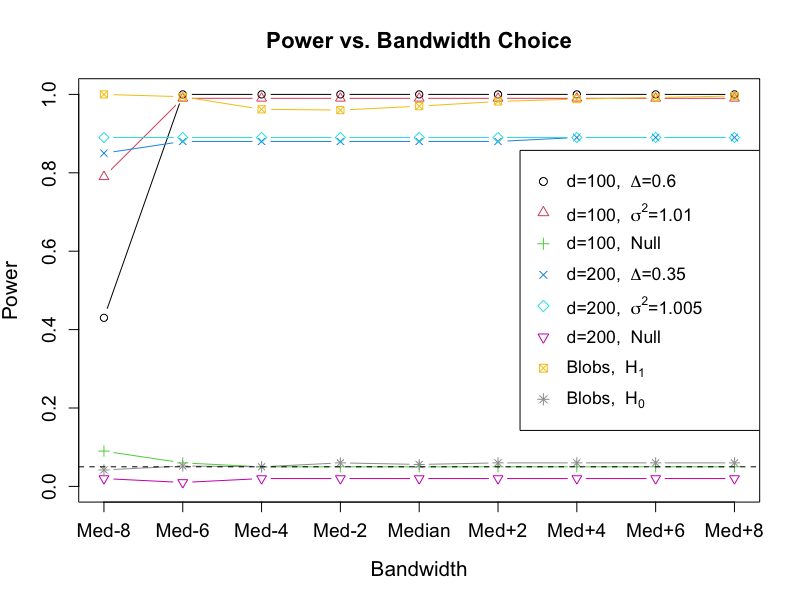}
	\caption{Estimated power based on 100 trials of the new test vs. bandwidth}
	\label{fig:band}
\end{figure}

We use 10,000 samples from each distribution and check the performance of the new test for each bandwidth choice under eight different settings at 0.05 significance level (Figure \ref{fig:band}). 

We see that the new test with the median heuristic not only controls the type I error well but also performs well even for the Blobs data compared to other choices of bandwidths. This result supports our conjecture in that the main data variation captured by the median heuristic reflects the difference between $P$ and $Q$ relatively well under the permutation null distribution. Hence, we use the median heuristic for the proposed test.

%%%%%%%%%%%%%%%%%%%%%%%%%%%%%%%%%%%%%%%%%%%%%%%%%%%%%%%%
%%%%%%%%%%%%%%%%%%%%%%%%%%%%%%%%%%%%%%%%%%%%%%%%%%%%%%%%%%%%%%%%%%%%%%%%%%%%%%%%%%%%%%%%%%%%%%%%%%%%%%%%%%%%%%%%

\section{Real data examples} \label{sec:real}

In this section, we illustrate the new test on two data sets: an age dataset and an eye movement dataset.

\subsection{Age dataset} 

\begin{figure}[h!]
	\centering
	\includegraphics[width=3in]{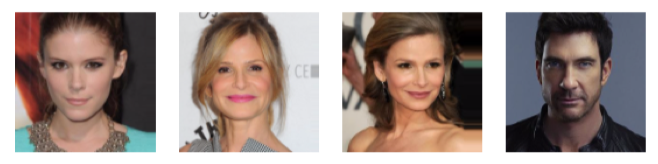}
	\caption{Sampling images from IMDb} 
	\label{fig:image}
\end{figure}

We now demonstrate the new test on real data using the age dataset from the IMDb-WIKI database \citep{rothe2018deep} (Figure \ref{fig:image}).  This dataset is publicly available at \url{https://data.vision.ee.ethz.ch/cvl/rrothe/imdb-wiki/}. It consists of 397,949 images of 19,545 celebrities with corresponding age labels, where the dimension is $d=4096$. Here, we follow the preprocessing of \cite{law2018bayesian}, and construct two groups according to the celebrity's age label. For example, the 10-15 group represents the images where the celebrity's age label is between 10 and 15. 

One problem for such large-scale datasets is that it is difficult to apply the existing kernel two-sample tests to determine whether the two samples are different or not. For example, the 15-20 age group consists of 16,282 images and the 20-25 age group has 39,957 images. Due to this large number of samples, it is computationally expensive for most existing tests. Table \ref{tab:imagetime} shows the computation time of the tests on 2.2 GHz Intel Core i7. We see that ME-full and SCF-full are too slow to obtain the test results. Although MMD-B is faster than ME-full and SCF-full, the test result would be unreliable due to lower power than ME-full and SCF-full (see Section \ref{sec:experiment}). On the other hand, the new test is still very fast with good performance.

\begin{table}[h!]
	\caption{\label{tab:imagetime}Runtimes in seconds for image dataset}
	\centering
	\begin{tabular}{c|cccc}
		\hline
		& New & ME-full & SCF-full & MMD-B  \\
		\hline
		15-20 vs 20-25    & 1.641& $>$3000 & $>$3000 & 48.19 \\
		\hline
	\end{tabular}
\end{table}

To illustrate how well the tests distinguish age groups, we conduct the testing procedures on subsets of the whole data so that ME-full and SCF-full are applicable. We randomly select 500 images from each group and repeat the experiment 500 times. Since this dataset has a lot of noise, we set the significance level to  0.001. 

The power of the tests is shown in Table \ref{tab:image}. Since this dataset has strong signals, all tests can easily distinguish between two age groups. However, the existing tests cannot capture the difference well in some cases. For example, SCF-full has 0 power in comparing the 30-35 and 35-40 age groups. MMD-B has low power in comparing 35-40 vs. 40-45 and 40-45 vs. 45-50. On the other hand, the new test consistently has high power in all these comparisons.

\begin{table}[h!]
	\caption{\label{tab:image}Estimated power of the tests for age dataset (randomly drawn subsets)}
	\centering
	\begin{tabular}{c|cccc}
		\hline
		Problem & New & ME-full & SCF-full & MMD-B   \\
		\hline
		15-20 vs 20-25    & 1.00 & 1.00 & 1.00 & 1.00 \\
		20-25 vs 25-30    & 1.00 & 1.00 & 1.00 & 0.80  \\
		25-30 vs 30-35    & 0.99 & 1.00 & 0.99  & 0.79 \\
		30-35 vs 35-40    & 1.00 & 1.00 & 0.00 &  0.83 \\
		35-40 vs 40-45    & 0.95 & 0.99 & 0.96 & 0.25  \\
		40-45 vs 45-50    & 0.93 & 0.73 & 0.85 & 0.40  \\
		\hline
	\end{tabular}
\end{table}

\subsection{Eye movement}

In this experiment, we examine how well the new test can distinguish samples of eye movement patterns \citep{salojarvi2005inferring}.    A subject was shown questions and asked to choose a sentence from a list  consisting of the correct answer (C), relevant sentences for the question (R), and irrelevant sentences for the question (I). In each case, 22 features are obtained, which are commonly used in psychological studies on eye movement. The dataset is publicly available at \url{https://www.openml.org/d/1044}.

Since this dataset consists of around 3000 samples (I: 3,804, R: 4,262, C:2,870) with low dimension, we focus more on the performance of the tests. We first conduct two-sample testing for the whole data to determine whether each group is different. Since only the new test can be applied to the unbalanced sample design, we draw 
the number of samples in the group with the smaller sample from the other group in each comparison. The results are shown in Table \ref{tab:eyefull}. We see that the new test and MMD-B reject all cases at 0.001 significance level. However, SCF-Full is not significant for the two comparisons, I vs. R and I vs. C, and also not significant for R vs. C at 0.001 significance level. ME-full is not significant for the comparison I vs. R.

We investigate the test statistics in more detail for this comparison. Table \ref{tab:eyefull} (last two columns) shows $p$-values of the new test statistics. We see that most $p$-values are almost zero, but $p$-value of $\bar{Z}_{D}$ for the case I vs. R is very large (0.460). This implies that the mean difference is mainly significant for the case I vs. R, while both the mean and the variance differences are significant for the other two cases. When the mean mainly differs (I vs. R), the new test captures this signal, while ME-full and SCF-full cannot capture this signal. MMD-B also captures this signal since MMD is very sensitive to the mean change (see argument in \cite{song2024generalized}). Indeed, \cite{salojarvi2005inferring} carried out linear discriminant analysis with  visualization to obtain a classification result and revealed that relevant and irrelevant sentences are relatively harder to separate than the other two cases.

\begin{table}[h!]
	\caption{\label{tab:eyefull}$p$-values for eye movement dataset}
	\centering
	\begin{tabular}{c|cccccc}
		\hline
		Problem & New & ME-full & SCF-full & MMD-B & $\bar{Z}_{W}$ & $\bar{Z}_{D}$  \\
		\hline
		I vs. R    & 0.000 & 0.197  & 0.516 & 0.000 & 0.000 & 0.460  \\
		I vs. C    & 0.000 & 0.000 & 0.800 & 0.000  & 0.000 & 0.000 \\
		R vs. C   & 0.000 & 0.000 & 0.013  & 0.000 & 0.000 & 0.000 \\
		\hline
	\end{tabular}
\end{table}

\begin{table}[h!]
	\caption{\label{tab:eye}Estimated power of the tests for eye movement dataset (randomly drawn subsets)}
	\centering
	\begin{tabular}{c|cccc}
		\hline
		\multicolumn{5}{c}{I vs. C} \\
		\hline
		$m=n$ & New & ME-full & SCF-full & MMD-B   \\
		\hline
		100    & 0.826 & 0.561 & 0.003 & 0.374 \\
		200    & 0.998 & 0.865 & 0.002 & 0.850  \\
		300    & 1.000 & 0.943 & 0.002 & 0.985 \\
		400    & 1.000 & 0.993  & 0.000 & 1.000 \\
		\hline
		\multicolumn{5}{c}{R vs. C} \\
		\hline
		$m=n$ & New & ME-full & SCF-full & MMD-B   \\
		\hline
		100    & 0.670 & 0.538 & 0.002 & 0.236 \\
		200    & 0.969 & 0.701 &0.000 & 0.685  \\
		300    & 0.999 & 0.987 & 0.000 & 0.941 \\
		400    & 1.000 & 1.000 & 0.002 & 0.988 \\
		\hline
	\end{tabular}
\end{table}
For the two cases where most tests are significant (I vs. C and R vs. C), we also examine the performance of the tests using subsets of the whole data. For each sample size, we generate 1,000 randomly selected subsets of the whole data and the significance level is set to  0.001 for all tests. The results are shown in Table \ref{tab:eye}. We see that SCF-full has no power and MMD-B shows lower power than the new test in all cases. ME-full exhibits higher power than SCF-full and MMD-B, but it is generally outperformed by the new test.

%%%%%%%%%%%%%%%%%%%%%%%%%%%%%%%%%%%%%%%%%%%%%%%%%%%%%%%%%%%%%%%%%%%%%%%%%%%%%%

\section{Discussion} \label{sec:discussion}

In this section, we briefly discuss some other blocking approaches along the same line as $b$, $B_{1}$, and $B_{2}$.
\begin{itemize}
	\item Approach 1 (A1): $B_{1} = \lfloor m/\sqrt{m}\rfloor$, $B_{2} = \lfloor n/\sqrt{n}\rfloor$, $b =\min(\lfloor m/B_{1}\rfloor, \lfloor n/B_{2}\rfloor)$.
	\item Approach 2 (A2): $B_{1} = \lfloor m/b\rfloor$, $B_{2} = \lfloor n/b\rfloor$, $b = \lfloor\sqrt{\min(m,n)}\rfloor$.
	\item Approach 3 (A3): $B_{1} = \lfloor m/b\rfloor$, $B_{2} = \lfloor n/b\rfloor$, $b = \lfloor\sqrt{\max(m,n)}\rfloor$.
\end{itemize}
Note that, under the balanced design, all approaches are equivalent to the block approach in \cite{zaremba2013b}.

To compare these three approaches to the new test, we check the performance of tests under the simulation setting used in Section \ref{subsec:simul2}. Here, the ratio of the two sample sizes $m$ and $n$ is fixed at 4:1 and 6:1.

\begin{table}[h!]
	\centering
	\caption{\label{tab:block1}Estimated power of the tests for each approach. $m = 4n$}
	\begin{tabular}{c|cccc|cccc}
		\hline
		$d=100$ & \multicolumn{4}{c|}{Sample sizes change $(a=0.03)$} & \multicolumn{4}{c}{Sample sizes change $(a=0.04)$} \\ 
		\hline
		Approach & A1 & A2 & A3  & New & A1 & A2 & A3  & New \\ 
		\hline
		$n = 500$ & 0.146 & 0.194 & \textbf{0.208} & \textbf{0.208} & 0.346 & 0.432 & \textbf{0.460} & 0.440 \\ 
		$n = 800$ & 0.264 & \textbf{0.374} & 0.360 & 0.360 & 0.494 & 0.634 & 0.620 & \textbf{0.646} \\
		$n = 1100$  &  0.410 & \textbf{0.532} &  0.502 &  0.514 & 0.710 & 0.792 & 0.796 & \textbf{0.806} \\
		$n = 1400$ & 0.530 & 0.648 & 0.622 & \textbf{0.656} &  0.836 & \textbf{0.916} & 0.906 & \textbf{0.916} \\
		$n = 1700$ & 0.640 & 0.764 & 0.758 & \textbf{0.768} & 0.902 & 0.950 & 0.944 & \textbf{0.958} \\
		\hline
	\end{tabular} 
	
	\vspace*{0.3cm}  
	
	\begin{tabular}{c|cccc|cccc}
		\hline
		$n=2000$ & \multicolumn{4}{c|}{Dimensions change $(a=0.01)$} & \multicolumn{4}{c}{Dimensions change $(a=0.02)$} \\ 
		\hline
		Approach & A1 & A2 & A3  & New & A1 & A2 & A3  & New \\ 
		\hline
		$d = 100$ & 0.166 & 0.218 & 0.218 & \textbf{0.224} & 0.282 & 0.370 & \textbf{0.378} & 0.364 \\ 
		$d = 200$ & 0.288 & 0.356 & 0.352 & \textbf{0.368} & 0.530 & 0.628 & \textbf{0.638} & 0.636\\
		$d = 300$  & 0.332 & \textbf{0.464} & 0.456 & 0.444 & 0.692 & \textbf{0.804} & \textbf{0.804} & \textbf{0.804} \\
		$d = 400$ & 0.438 & \textbf{0.574} & 0.548 & 0.564 & 0.790 & \textbf{0.892} & 0.880 & \textbf{0.892}  \\
		$d = 500$ & 0.574 & 0.688 & 0.678 & \textbf{0.700} & 0.880 & \textbf{0.942} & 0.940 & \textbf{0.942}  \\
		\hline
	\end{tabular} 
\end{table}

The results are shown in Tables \ref{tab:block1} and \ref{tab:block2}. When $m = 4n$, we see that A2, A3 and the new test perform well when the sample size is small. On the other hand, the new test dominates in power when the sample size is large. When the dimension is low, A3 and the new test perform well, while A2 and the new test exhibit high power when the dimension is high. When $m = 6n$, A2 exhibits high power when the sample size is large, while the new approach  generally show good performance when the dimension is high. Therefore, we still recommend the proposed block approach in general scenarios.

\begin{table} [h!]
	\centering
	\caption{\label{tab:block2}Estimated power of the tests for each approach. $m = 6n$}
	\begin{tabular}{c|cccc|cccc}
		\hline
		$d=100$ & \multicolumn{4}{c|}{Sample sizes change $(a=0.03)$} & \multicolumn{4}{c}{Sample sizes change $(a=0.04)$} \\
		\hline
		Approach & A1 & A2 & A3  & New & A1 & A2 & A3  & New \\
		\hline
		$n = 400$ & 0.122 & 0.188 & 0.165 & \textbf{0.198} & 0.271 & \textbf{0.384} & 0.362 & 0.367 \\ 
		$n = 600$ & 0.220 & \textbf{0.303} & 0.299 & 0.295 & 0.428 & 0.547 & 0.527 & \textbf{0.548} \\
		$n = 800$  &  0.295 & 0.392 &  \textbf{0.402} &  0.398 & 0.549 & \textbf{0.697} & 0.671 & 0.667 \\
		$n = 1000$ & 0.400 & \textbf{0.507} & 0.504 & 0.487 &  0.670 & \textbf{0.796} & 0.766 & 0.754 \\
		$n = 1200$ & 0.483 & \textbf{0.610} & 0.588 & 0.600 & 0.775 & \textbf{0.880} & 0.869 & 0.872 \\
		\hline
	\end{tabular} 
	
	\vspace*{0.3cm}  
	
	\begin{tabular}{c|cccc|cccc}
		\hline
		$n=1400$ & \multicolumn{4}{c|}{Dimensions change $(a=0.01)$} & \multicolumn{4}{c}{Dimensions change $(a=0.02)$} \\ 
		\hline
		Approach & A1 & A2 & A3  & New & A1 & A2 & A3  & New \\
		\hline
		$d = 100$ & 0.146 & 0.186 & \textbf{0.212} & 0.196 & 0.210 & 0.282 & 0.276 & \textbf{0.296} \\ 
		$d = 200$ & 0.258 & 0.354 & 0.362 & \textbf{0.364} & 0.402 & \textbf{0.540} & 0.518 & \textbf{0.540} \\
		$d = 300$  & 0.388 & 0.488 & \textbf{0.522} & 0.516 & 0.514 & 0.642 & 0.644 & \textbf{0.648} \\
		$d = 400$ & 0.458 & 0.548 & 0.558 & \textbf{0.576} & 0.628 & 0.724 & 0.742 & \textbf{0.778}  \\
		$d = 500$ & 0.574 & 0.692 & 0.698 & \textbf{0.706} &0.738 & 0.834 & \textbf{0.846} & 0.844  \\
		\hline
	\end{tabular} 
\end{table}

%%%%%%%%%%%%%%%%%%%%%%%%%%%%%%%%%%%%%%%%%%%%%%%%%%%%%%%%%%%%%%%%%%%%%%%%%%%%%%%%%%%%%%%%%%%%%%%%%%%

\section{Conclusion} \label{sec:conclusion}

In this paper, we proposed a new kernel two-sample test for large-scale data. The new test combines the strengths of two statistics, enabling it to effectively detect differences in distributions. The asymptotic distributions of the test statistics facilitate its application to large datasets. The new approach is robust to high dimensions, applicable to  unbalanced sample designs, and  does not require parameter tuning procedures through data splitting, making it computationally efficient.

\bmhead{Acknowledgements}

Hoseung Song is supported by the National Research Foundation of Korea (NRF) grant funded by the Korea government (MSIT) RS-2025-16066571. Hao Chen is supported in part by the NSF award DMS-1848579 and DMS-2311399. 

%%%%%%%%%%%%%%%%%%%%%%%%%%%%%%%%%%%%%%%%%%%%%%%%%%%%%%%%

\bibliography{sn-bibliography}

\begin{thebibliography}{}
\providecommand{\doi}[1]{\url{https://doi.org/#1}}
\bibcommenthead

\bibitem[\protect\citeauthoryear{Bai and Saranadasa}{Bai and
  Saranadasa}{1996}]{bai1996effect}
Bai, Z. and H.~Saranadasa. 1996.
\newblock Effect of high dimension: by an example of a two sample problem.
\newblock {\em Statistica Sinica\/}: 311--329 .

\bibitem[\protect\citeauthoryear{Baumgartner, Wei{\ss}, and
  Schindler}{Baumgartner et~al.}{1998}]{baumgartner1998nonparametric}
Baumgartner, W., P.~Wei{\ss}, and H.~Schindler. 1998.
\newblock A nonparametric test for the general two-sample problem.
\newblock {\em Biometrics\/}: 1129--1135 .

\bibitem[\protect\citeauthoryear{Biswas and Ghosh}{Biswas and
  Ghosh}{2014}]{biswas2014nonparametric}
Biswas, M. and A.K. Ghosh. 2014.
\newblock A nonparametric two-sample test applicable to high dimensional data.
\newblock {\em Journal of Multivariate Analysis\/}~123: 160--171 .

\bibitem[\protect\citeauthoryear{Carlini and Wagner}{Carlini and
  Wagner}{2017}]{carlini2017adversarial}
Carlini, N. and D.~Wagner 2017.
\newblock Adversarial examples are not easily detected: Bypassing ten detection
  methods.
\newblock In {\em Proceedings of the 10th ACM workshop on artificial
  intelligence and security}, pp.\  3--14.

\bibitem[\protect\citeauthoryear{Chen, Chen, and Su}{Chen
  et~al.}{2018}]{chen2018weighted}
Chen, H., X.~Chen, and Y.~Su. 2018.
\newblock A weighted edge-count two-sample test for multivariate and object
  data.
\newblock {\em Journal of the American Statistical Association\/}~{\em
  113\/}(523): 1146--1155 .

\bibitem[\protect\citeauthoryear{Chen and Friedman}{Chen and
  Friedman}{2017}]{chen2017new}
Chen, H. and J.H. Friedman. 2017.
\newblock A new graph-based two-sample test for multivariate and object data.
\newblock {\em Journal of the American statistical association\/}~{\em
  112\/}(517): 397--409 .

\bibitem[\protect\citeauthoryear{Chen and Zhang}{Chen and
  Zhang}{2013}]{chen2013graph}
Chen, H. and N.R. Zhang. 2013.
\newblock Graph-based tests for two-sample comparisons of categorical data.
\newblock {\em Statistica Sinica\/}: 1479--1503 .

\bibitem[\protect\citeauthoryear{Chwialkowski, Ramdas, Sejdinovic, and
  Gretton}{Chwialkowski et~al.}{2015}]{chwialkowski2015fast}
Chwialkowski, K.P., A.~Ramdas, D.~Sejdinovic, and A.~Gretton 2015.
\newblock Fast two-sample testing with analytic representations of probability
  measures.
\newblock In {\em Advances in Neural Information Processing Systems}, pp.\
  1981--1989.

\bibitem[\protect\citeauthoryear{Fox and Dimmic}{Fox and
  Dimmic}{2006}]{fox2006two}
Fox, R.J. and M.W. Dimmic. 2006.
\newblock A two-sample bayesian t-test for microarray data.
\newblock {\em BMC bioinformatics\/}~{\em 7\/}(1): 1--11 .

\bibitem[\protect\citeauthoryear{Friedman and Rafsky}{Friedman and
  Rafsky}{1979}]{friedman1979multivariate}
Friedman, J.H. and L.C. Rafsky. 1979.
\newblock Multivariate generalizations of the wald-wolfowitz and smirnov
  two-sample tests.
\newblock {\em The Annals of Statistics\/}: 697--717 .

\bibitem[\protect\citeauthoryear{Gao, Liu, Zhang, Han, Liu, Niu, and
  Sugiyama}{Gao et~al.}{2020}]{gao2020maximum}
Gao, R., F.~Liu, J.~Zhang, B.~Han, T.~Liu, G.~Niu, and M.~Sugiyama. 2020.
\newblock Maximum mean discrepancy is aware of adversarial attacks.
\newblock {\em arXiv preprint arXiv:2010.11415\/} .

\bibitem[\protect\citeauthoryear{Gretton et~al.}{Gretton
  et~al.}{2012}]{gretton2012kernel}
Gretton, A. et~al. 2012.
\newblock A kernel two-sample test.
\newblock {\em Journal of Machine Learning Research\/}~{\em 13\/}(Mar):
  723--773 .

\bibitem[\protect\citeauthoryear{Gretton, Borgwardt, Rasch, Sch{\"o}lkopf, and
  Smola}{Gretton et~al.}{2007}]{gretton2007kernel}
Gretton, A., K.M. Borgwardt, M.~Rasch, B.~Sch{\"o}lkopf, and A.J. Smola 2007.
\newblock A kernel method for the two-sample-problem.
\newblock In {\em Advances in neural information processing systems}, pp.\
  513--520.

\bibitem[\protect\citeauthoryear{Gretton, Fukumizu, Harchaoui, and
  Sriperumbudur}{Gretton et~al.}{2009}]{gretton2009fast}
Gretton, A., K.~Fukumizu, Z.~Harchaoui, and B.K. Sriperumbudur 2009.
\newblock A fast, consistent kernel two-sample test.
\newblock In {\em Advances in neural information processing systems}, pp.\
  673--681.

\bibitem[\protect\citeauthoryear{Gretton, Sejdinovic, Strathmann, Balakrishnan,
  Pontil, Fukumizu, and Sriperumbudur}{Gretton
  et~al.}{2012}]{gretton2012optimal}
Gretton, A., D.~Sejdinovic, H.~Strathmann, S.~Balakrishnan, M.~Pontil,
  K.~Fukumizu, and B.K. Sriperumbudur. 2012.
\newblock Optimal kernel choice for large-scale two-sample tests.
\newblock {\em Advances in neural information processing systems\/}~25 .

\bibitem[\protect\citeauthoryear{Grosse, Manoharan, Papernot, Backes, and
  McDaniel}{Grosse et~al.}{2017}]{grosse2017statistical}
Grosse, K., P.~Manoharan, N.~Papernot, M.~Backes, and P.~McDaniel. 2017.
\newblock On the (statistical) detection of adversarial examples.
\newblock {\em arXiv preprint arXiv:1702.06280\/} .

\bibitem[\protect\citeauthoryear{Grossfield and Zuckerman}{Grossfield and
  Zuckerman}{2009}]{grossfield2009quantifying}
Grossfield, A. and D.M. Zuckerman. 2009.
\newblock Quantifying uncertainty and sampling quality in biomolecular
  simulations.
\newblock {\em Annual reports in computational chemistry\/}~5: 23--48 .

\bibitem[\protect\citeauthoryear{Hediger, Michel, and N{\"a}f}{Hediger
  et~al.}{2019}]{hediger2019use}
Hediger, S., L.~Michel, and J.~N{\"a}f. 2019.
\newblock On the use of random forest for two-sample testing.
\newblock {\em arXiv preprint arXiv:1903.06287\/} .

\bibitem[\protect\citeauthoryear{Hettmansperger, M{\"o}tt{\"o}nen, and
  Oja}{Hettmansperger et~al.}{1998}]{hettmansperger1998affine}
Hettmansperger, T.P., J.~M{\"o}tt{\"o}nen, and H.~Oja. 1998.
\newblock Affine invariant multivariate rank tests for several samples.
\newblock {\em Statistica Sinica\/}: 785--800 .

\bibitem[\protect\citeauthoryear{Ho and Shieh}{Ho and Shieh}{2006}]{ho2006two}
Ho, H.c. and G.S. Shieh. 2006.
\newblock Two-stage u-statistics for hypothesis testing.
\newblock {\em Scandinavian journal of statistics\/}~{\em 33\/}(4): 861--873 .

\bibitem[\protect\citeauthoryear{Jayasumana, Ramalingam, Veit, Glasner,
  Chakrabarti, and Kumar}{Jayasumana et~al.}{2024}]{jayasumana2024rethinking}
Jayasumana, S., S.~Ramalingam, A.~Veit, D.~Glasner, A.~Chakrabarti, and
  S.~Kumar 2024.
\newblock Rethinking fid: Towards a better evaluation metric for image
  generation.
\newblock In {\em Proceedings of the IEEE/CVF Conference on Computer Vision and
  Pattern Recognition}, pp.\  9307--9315.

\bibitem[\protect\citeauthoryear{Jitkrittum, Szab{\'o}, Chwialkowski, and
  Gretton}{Jitkrittum et~al.}{2016}]{jitkrittum2016interpretable}
Jitkrittum, W., Z.~Szab{\'o}, K.P. Chwialkowski, and A.~Gretton. 2016.
\newblock Interpretable distribution features with maximum testing power.
\newblock {\em Advances in Neural Information Processing Systems\/}~29:
  181--189 .

\bibitem[\protect\citeauthoryear{Kent~IV, Muller, Anderson, Goddard~III, and
  Feldmann}{Kent~IV et~al.}{2007}]{kent2007efficient}
Kent~IV, D.R., R.P. Muller, A.G. Anderson, W.A. Goddard~III, and M.T. Feldmann.
  2007.
\newblock Efficient algorithm for “on-the-fly” error analysis of local or
  distributed serially correlated data.
\newblock {\em Journal of computational chemistry\/}~{\em 28\/}(14): 2309--2316
  .

\bibitem[\protect\citeauthoryear{Kirchler, Khorasani, Kloft, and
  Lippert}{Kirchler et~al.}{2020}]{kirchler2020two}
Kirchler, M., S.~Khorasani, M.~Kloft, and C.~Lippert 2020.
\newblock Two-sample testing using deep learning.
\newblock In {\em International Conference on Artificial Intelligence and
  Statistics}, pp.\  1387--1398. PMLR.

\bibitem[\protect\citeauthoryear{Kohout and Pevn{\`y}}{Kohout and
  Pevn{\`y}}{2017}]{kohout2017network}
Kohout, J. and T.~Pevn{\`y}. 2017.
\newblock Network traffic fingerprinting based on approximated kernel
  two-sample test.
\newblock {\em IEEE Transactions on Information Forensics and Security\/}~{\em
  13\/}(3): 788--801 .

\bibitem[\protect\citeauthoryear{Kolmogorov}{Kolmogorov}{1933}]{kolmogorov1933sulla}
Kolmogorov, A. 1933.
\newblock Sulla determinazione emp{\'\i}rica di uma legge di distribuzione .

\bibitem[\protect\citeauthoryear{Law, Sutherland, Sejdinovic, and Flaxman}{Law
  et~al.}{2018}]{law2018bayesian}
Law, H.C.L., D.~Sutherland, D.~Sejdinovic, and S.~Flaxman 2018.
\newblock Bayesian approaches to distribution regression.
\newblock In {\em International Conference on Artificial Intelligence and
  Statistics}, pp.\  1167--1176.

\bibitem[\protect\citeauthoryear{Liu, Xu, Lu, Zhang, Gretton, and
  Sutherland}{Liu et~al.}{2020}]{liu2020learning}
Liu, F., W.~Xu, J.~Lu, G.~Zhang, A.~Gretton, and D.J. Sutherland 2020.
\newblock Learning deep kernels for non-parametric two-sample tests.
\newblock In {\em International Conference on Machine Learning}, pp.\
  6316--6326. PMLR.

\bibitem[\protect\citeauthoryear{Lopez-Paz and Oquab}{Lopez-Paz and
  Oquab}{2016}]{lopez2016revisiting}
Lopez-Paz, D. and M.~Oquab. 2016.
\newblock Revisiting classifier two-sample tests.
\newblock {\em arXiv preprint arXiv:1610.06545\/} .

\bibitem[\protect\citeauthoryear{Mann and Whitney}{Mann and
  Whitney}{1947}]{mann1947test}
Mann, H.B. and D.R. Whitney. 1947.
\newblock On a test of whether one of two random variables is stochastically
  larger than the other.
\newblock {\em The annals of mathematical statistics\/}: 50--60 .

\bibitem[\protect\citeauthoryear{Meng-Papaxanthos, Zhang, Li, Cuturi, Noble,
  and Vert}{Meng-Papaxanthos et~al.}{2023}]{meng2023lsmmd}
Meng-Papaxanthos, L., R.~Zhang, G.~Li, M.~Cuturi, W.S. Noble, and J.P. Vert.
  2023.
\newblock Lsmmd-ma: scaling multimodal data integration for single-cell
  genomics data analysis.
\newblock {\em Bioinformatics\/}~{\em 39\/}(7): btad420 .

\bibitem[\protect\citeauthoryear{Oja}{Oja}{2010}]{oja2010multivariate}
Oja, H. 2010.
\newblock {\em Multivariate nonparametric methods with R: an approach based on
  spatial signs and ranks}.
\newblock Springer Science \& Business Media.

\bibitem[\protect\citeauthoryear{Osborne, Patrangenaru, Ellingson, Groisser,
  and Schwartzman}{Osborne et~al.}{2013}]{osborne2013nonparametric}
Osborne, D., V.~Patrangenaru, L.~Ellingson, D.~Groisser, and A.~Schwartzman.
  2013.
\newblock Nonparametric two-sample tests on homogeneous riemannian manifolds,
  cholesky decompositions and diffusion tensor image analysis.
\newblock {\em Journal of Multivariate Analysis\/}~119: 163--175 .

\bibitem[\protect\citeauthoryear{Ozier-Lafontaine, Fourneaux, Durif, Arsenteva,
  Vallot, Gandrillon, Gonin-Giraud, Michel, and Picard}{Ozier-Lafontaine
  et~al.}{2024}]{ozier2024kernel}
Ozier-Lafontaine, A., C.~Fourneaux, G.~Durif, P.~Arsenteva, C.~Vallot,
  O.~Gandrillon, S.~Gonin-Giraud, B.~Michel, and F.~Picard. 2024.
\newblock Kernel-based testing for single-cell differential analysis.
\newblock {\em Genome Biology\/}~{\em 25\/}(1): 114 .

\bibitem[\protect\citeauthoryear{Pan, Tian, Wang, and Zhang}{Pan
  et~al.}{2018}]{pan2018ball}
Pan, W., Y.~Tian, X.~Wang, and H.~Zhang. 2018.
\newblock Ball divergence: nonparametric two sample test.
\newblock {\em Annals of statistics\/}~{\em 46\/}(3): 1109 .

\bibitem[\protect\citeauthoryear{Ramdas, Reddi, Poczos, Singh, and
  Wasserman}{Ramdas et~al.}{2015}]{ramdas2015adaptivity}
Ramdas, A., S.J. Reddi, B.~Poczos, A.~Singh, and L.~Wasserman. 2015.
\newblock Adaptivity and computation-statistics tradeoffs for kernel and
  distance based high dimensional two sample testing.
\newblock {\em arXiv preprint arXiv:1508.00655\/} .

\bibitem[\protect\citeauthoryear{Rosenbaum}{Rosenbaum}{2005}]{rosenbaum2005exact}
Rosenbaum, P.R. 2005.
\newblock An exact distribution-free test comparing two multivariate
  distributions based on adjacency.
\newblock {\em Journal of the Royal Statistical Society: Series B (Statistical
  Methodology)\/}~{\em 67\/}(4): 515--530 .

\bibitem[\protect\citeauthoryear{Rothe, Timofte, and Van~Gool}{Rothe
  et~al.}{2018}]{rothe2018deep}
Rothe, R., R.~Timofte, and L.~Van~Gool. 2018.
\newblock Deep expectation of real and apparent age from a single image without
  facial landmarks.
\newblock {\em International Journal of Computer Vision\/}~{\em 126\/}(2-4):
  144--157 .

\bibitem[\protect\citeauthoryear{Rousson}{Rousson}{2002}]{rousson2002distribution}
Rousson, V. 2002.
\newblock On distribution-free tests for the multivariate two-sample
  location-scale model.
\newblock {\em Journal of multivariate analysis\/}~{\em 80\/}(1): 43--57 .

\bibitem[\protect\citeauthoryear{Saloj{\"a}rvi, Puolam{\"a}ki, Simola, Kovanen,
  Kojo, and Kaski}{Saloj{\"a}rvi et~al.}{2005}]{salojarvi2005inferring}
Saloj{\"a}rvi, J., K.~Puolam{\"a}ki, J.~Simola, L.~Kovanen, I.~Kojo, and
  S.~Kaski 2005.
\newblock Inferring relevance from eye movements: Feature extraction.
\newblock In {\em Workshop at NIPS 2005, in Whistler, BC, Canada, on December
  10, 2005.}, pp.\ ~45.

\bibitem[\protect\citeauthoryear{Schilling}{Schilling}{1986}]{schilling1986multivariate}
Schilling, M.F. 1986.
\newblock Multivariate two-sample tests based on nearest neighbors.
\newblock {\em Journal of the American Statistical Association\/}~{\em
  81\/}(395): 799--806 .

\bibitem[\protect\citeauthoryear{Schott}{Schott}{2007}]{schott2007test}
Schott, J.R. 2007.
\newblock A test for the equality of covariance matrices when the dimension is
  large relative to the sample sizes.
\newblock {\em Computational Statistics \& Data Analysis\/}~{\em 51\/}(12):
  6535--6542 .

\bibitem[\protect\citeauthoryear{Song and Chen}{Song and
  Chen}{2024}]{song2024generalized}
Song, H. and H.~Chen. 2024.
\newblock Generalized kernel two-sample tests.
\newblock {\em Biometrika\/}~{\em 111\/}(3): 755--770 .

\bibitem[\protect\citeauthoryear{Sriperumbudur, Gretton, Fukumizu,
  Sch{\"o}lkopf, and Lanckriet}{Sriperumbudur
  et~al.}{2010}]{sriperumbudur2010hilbert}
Sriperumbudur, B.K., A.~Gretton, K.~Fukumizu, B.~Sch{\"o}lkopf, and G.R.
  Lanckriet. 2010.
\newblock Hilbert space embeddings and metrics on probability measures.
\newblock {\em Journal of Machine Learning Research\/}~{\em 11\/}(Apr):
  1517--1561 .

\bibitem[\protect\citeauthoryear{Srivastava and Du}{Srivastava and
  Du}{2008}]{srivastava2008test}
Srivastava, M.S. and M.~Du. 2008.
\newblock A test for the mean vector with fewer observations than the
  dimension.
\newblock {\em Journal of Multivariate Analysis\/}~{\em 99\/}(3): 386--402 .

\bibitem[\protect\citeauthoryear{Srivastava and Yanagihara}{Srivastava and
  Yanagihara}{2010}]{srivastava2010testing}
Srivastava, M.S. and H.~Yanagihara. 2010.
\newblock Testing the equality of several covariance matrices with fewer
  observations than the dimension.
\newblock {\em Journal of Multivariate Analysis\/}~{\em 101\/}(6): 1319--1329 .

\bibitem[\protect\citeauthoryear{Sutherland, Tung, Strathmann, De, Ramdas,
  Smola, and Gretton}{Sutherland et~al.}{2016}]{sutherland2016generative}
Sutherland, D.J., H.Y. Tung, H.~Strathmann, S.~De, A.~Ramdas, A.~Smola, and
  A.~Gretton. 2016.
\newblock Generative models and model criticism via optimized maximum mean
  discrepancy.
\newblock {\em arXiv preprint arXiv:1611.04488\/} .

\bibitem[\protect\citeauthoryear{Sutherland, Arbel, and Gretton}{Sutherland
  et~al.}{2018}]{sutherland2018demystifying}
Sutherland, J., M.~Arbel, and A.~Gretton 2018.
\newblock Demystifying mmd gans.
\newblock In {\em International conference for learning representations},
  Volume~6.

\bibitem[\protect\citeauthoryear{Sz{\'e}kely and Rizzo}{Sz{\'e}kely and
  Rizzo}{2013}]{szekely2013energy}
Sz{\'e}kely, G.J. and M.L. Rizzo. 2013.
\newblock Energy statistics: A class of statistics based on distances.
\newblock {\em Journal of statistical planning and inference\/}~{\em 143\/}(8):
  1249--1272 .

\bibitem[\protect\citeauthoryear{Wald and Wolfowitz}{Wald and
  Wolfowitz}{1940}]{wald1940test}
Wald, A. and J.~Wolfowitz. 1940.
\newblock On a test wether two samples are from the same distribution.
\newblock {\em Ann. Math. Stat\/}~11: 147--162 .

\bibitem[\protect\citeauthoryear{Wynne and Duncan}{Wynne and
  Duncan}{2020}]{wynne2020kernel}
Wynne, G. and A.B. Duncan. 2020.
\newblock A kernel two-sample test for functional data.
\newblock {\em arXiv preprint arXiv:2008.11095\/} .

\bibitem[\protect\citeauthoryear{Zaremba, Gretton, and Blaschko}{Zaremba
  et~al.}{2013}]{zaremba2013b}
Zaremba, W., A.~Gretton, and M.~Blaschko 2013.
\newblock B-test: A non-parametric, low variance kernel two-sample test.
\newblock In {\em Advances in neural information processing systems}, pp.\
  755--763.

\bibitem[\protect\citeauthoryear{Zhang and Chen}{Zhang and
  Chen}{2017}]{zhang2017graph}
Zhang, J. and H.~Chen. 2017.
\newblock Graph-based two-sample tests for data with repeated observations.
\newblock {\em arXiv preprint arXiv:1711.04349\/} .

\end{thebibliography}

%%%%%%%%%%%%%%%%%%%%%%%%%%%%%%%%%%%%%%%%%%%%%%%%%%%%%%%%

\end{document}